\documentclass[12pt]{article}
\usepackage{color}
\usepackage{a4}
\usepackage{latexsym}
\usepackage{cite}
\usepackage{bm}
\usepackage{lineno}
\usepackage{epsfig}
\usepackage{graphicx}

\usepackage{pslatex}
\usepackage[latin1]{inputenc}
\usepackage[T1]{fontenc}

\newcommand{\lsim}{\raisebox{-0.7mm}{$\;\stackrel{<}{{\scriptstyle
 \sim}}\: $} }

\newcommand{\beq}{\begin{equation}}
\newcommand{\eeq}{\end{equation}}
\newcommand{\bea}{\begin{eqnarray}}
\newcommand{\eea}{\end{eqnarray}}
\newcommand{\nn}{\nonumber}
\newcommand{\MSb}{$\overline{\mbox{MS}}$}
\newcommand{\ra}{\rightarrow}

\newcommand{\als}{\alpha_{\rm s}}
\newcommand{\ars}{a_{\rm s}}

\newcommand{\muf}{\mu_{\:\!\!f}^{}}
\newcommand{\mufs}{\mu_{\:\!\!f}^{\,2}}

\newcommand{\hspn}{{\hspace{-6mm}}}
\newcommand{\hspp}{{\hspace{5mm}}}

\def\frct#1#2{\mbox{\footnotesize{$\displaystyle\frac{#1}{#2}$}}}

\def\as(#1){{\alpha_{\rm s}^{\,#1}}}
\def\ar(#1){{a_{\rm s}^{\,#1}}}

\def\gns(#1){{\gamma_{\,\rm ns}^{\,(#1)}}}

\def\z#1{{\zeta_{#1}^{}}}
\def\zts{{\zeta_{3}^{\,2}}}

\def\B(#1,#2){{\beta_{#1}^{\,#2}}}

\def\nc{{n_c}}
\def\ncs{{n_{c}^{\,2}}}
\def\ca{{C^{}_A}}
\def\cas{{C^{\,2}_A}}
\def\cat{{C^{\,3}_A}}
\def\caf{{C^{\,4}_A}}
\def\cf{{C^{}_F}}
\def\cfs{{C^{\, 2}_F}}
\def\cft{{C^{\, 3}_F}}
\def\cff{{C^{\, 4}_F}}
\def\cfi{{C^{\, 5}_F}}
\def\nf{{n^{}_{\! f}}}

\def\nfs{{n^{\,2}_{\! f}}}
\def\nft{{n^{\,3}_{\! f}}}
\def\nff{{n^{\,4}_{\! f}}}

\def\dtFFnc{{\frct{d_{abc}d^{\,abc}}{N_F }}}
\def\dfAAna{{\frct{d^{\,(4)}_{AA}}{N_A }}}

\def\dfFAnc{{\frct{d^{\,(4)}_{FA}}{N_F }}}
\def\dfFFnc{{\frct{d^{\,(4)}_{FF}}{N_F }}}

\def\xm1{{(1 \! - \! x)}}

\begin{document}
\setlength{\parskip}{0.2cm}
\setlength{\baselineskip}{0.55cm}

\begin{titlepage}
\noindent
Nikhef 2018-066 \hfill December 2018\\
DESY 18--227 \\
LTH 1191 \\
\vspace{0.6cm}
%
%
\begin{center}
{\LARGE \bf Five-loop contributions to low-${\bm N}$\\[1ex]
non-singlet anomalous dimensions in QCD}\\ 
\vspace{1.4cm}
\large
F. Herzog$^{\, a}$, S. Moch$^{\, b}$, B. Ruijl$^{\, c}$, T. Ueda$^{\, d}$, 
J.A.M. Vermaseren$^{\, a}$ and A. Vogt$^{\, e}$\\
\vspace{1.2cm}
\normalsize
{\it $^a$Nikhef Theory Group \\
\vspace{0.5mm}
Science Park 105, 1098 XG Amsterdam, The Netherlands} \\
\vspace{4mm}
{\it $^b$II.~Institute for Theoretical Physics, Hamburg University\\
Luruper Chaussee 149, D-22761 Hamburg, Germany}\\
\vspace{4mm}
{\it $^c$Institute for Theoretical Physics, ETH Z\"urich\\
\vspace{0.5mm}
Wolfgang-Pauli-Str.~27, 8093 Z\"urich, Switzerland}\\
\vspace{4mm}
{\it $^d$ Department of Materials and Life Science, Seikei University\\
\vspace{0.5mm}
3-3-1 Kichijoji Kitamachi, Musashino-shi, Tokyo 180-8633, Japan}\\
\vspace{4mm}
{\it $^e$Department of Mathematical Sciences, University of Liverpool\\
\vspace{0.5mm}
Liverpool L69 3BX, United Kingdom}\\
\vspace{1.4cm}
{\large \bf Abstract}
\vspace{-0.2cm}
\end{center}
We present the first calculations of 
next-to-next-to-next-to-next-to-leading order (N$^4$LO) 
contributions to anomalous dimensions of spin-$N$ twist-2 operators
in perturbative QCD. 
Specifically, we have obtained the respective non-singlet 
quark-quark anomalous dimensions at $N=2$ and $N=3$ to the fifth 
order in the 
strong coupling $\als$. 
These results set the scale for the N$^4$LO contributions to the 
evolution of the non-singlet quark distributions of hadrons outside 
the small-$x$ region, and facilitate a first approximate 
determination of the five-loop cusp anomalous dimension.
While the N$^4$LO coefficients are larger than expected from the 
lower-order results, their inclusion stabilizes the perturbative
expansions for three or more light flavours at a sub-percent 
accuracy for $\als < 0.3$.

\vspace*{0.2cm}
\end{titlepage}


\noindent
The anomalous dimensions $\gamma_{\,\rm ik}^{}(N)$ of spin-$N$ twist-2 
operators are important quantities in perturbative QCD. 
They are closely related, by an integer-$N$ Mellin transform, to the
splitting functions $P_{\,\rm ik}(x)$ that govern the scale dependence 
(evolution) of the parton densities of hadrons, and are hence directly 
relevant to the analysis of hard processes at the LHC.
The coefficients $A_{\rm k}$ of the leading large-$N$ term of the 
diagonal ($\rm i = k$) anomalous dimensions in the standard \MSb\
scheme \cite{Korch89,ABall00,DMS05}
\beq
\label{Ntoinf}
  \gamma_{\,\rm kk}^{}(N) \; = \;
    A_{\rm k} \, \ln \widetilde{N} 
  \,-\, B_{\rm k}
  \,+\, C_{\rm k} \: N^{\,-1} \ln \widetilde{N}
  \,-\, \big( D_{\rm k} - \frct{1}{2}\: A_{\rm k} \big) \, N^{-1}
  \,+\, {\cal O} \big( N^{\,-2} \ln^{\,n} \widetilde{N} \big)
\eeq
(with $\ln \widetilde{N} = \ln N + \gamma_{\:\!e}$, where 
$\gamma_{\:\!e}$ is the Euler-Mascheroni constant) are identical to 
the (lightlike) cusp anomalous dimensions \cite{Korch89}, and thus
relevant well beyond the evolution of parton distributions.

At present, the splitting functions are fully known to three loops
(next-to-next-to-leading order, N$^2$LO), see refs.~\cite{MVV3,MVV4} 
for the main unpolarized case and refs.~\cite{MVVp1,MVVp2} for the 
helicity-dependent case. At four loops, the non-singlet quark-quark
splitting functions have been determined analytically in the limit of 
a large number of colours $\nc$; the remaining terms are known with 
a high numerical accuracy except at momentum fractions 
$x \lsim 10^{\,-2}$ \cite{MRUVV1}. 
The less advanced present status in the flavour-singlet sector beyond 
the leading large-$\nf$ terms \cite{DRUVV} has been summarized in 
ref.~\cite{avLL18}. 

In this letter, we report on the first complete calculations of 
five-loop (N$^4$LO) twist-2 anomalous dimensions in QCD and its
generalization to a general gauge group.
Specifically, we have computed $\gamma_{\,\rm ns}^{\:+}(N\!=\!2)$ and  
$\gamma_{\,\rm ns}^{\;k}(N\!=\!3)$ for $k = -, \rm v$. 
The superscripts refer to the combinations of quark densities
\beq
\label{qns}
   q^{\,\pm}_{ab} \;=\; q_{a}^{} \pm \bar{q}_{a}^{}
                      - ( q_{b}^{} \pm \bar{q}_{b}^{} )
\; , \quad
   q_{\rm v}^{} \;=\; {\textstyle \sum_{\,a=1}^{\,\nf}} 
   ( q_{a}^{} - \bar{q}_{a}^{} )
\eeq
and $\nf$ represents the number of effectively massless flavours.
These results set the scale for the N$^4$LO corrections to the
evolution of the non-singlet quark distributions outside the 
small-$x$ region.  In particular they allow, together with 
$\gamma_{\,\rm ns}^{\,-,\rm v}(N\!=\!1) = 0$ and specific properties 
in the large-$\nc$ limit, see below, first serious (if unavoidably 
rough) estimates of the five-loop cusp anomalous dimension.

In terms of operator definitions and renormalization, the present 
calculation is a direct generalization of ref.~\cite{MRUVV1}. 
The computation of the required five-loop self-energy integrals is
performed as in refs.~\cite{beta4,Hgg4}, i.e., we employ a recent
implementation \cite{HR-Rstar,brRC17} on the local R$^\ast$ operation 
\cite{Rstar1,Rstar2,Rstar3} to reduce these to four-loop integrals
that can be evaluated by the {\sc Forcer} program \cite{Forcer}.
All our symbolic manipulations are carried out using the latest
version \cite{FORM42} of {\sc Form} \cite{FORM3,FORM4}. 
The five-loop computation of $\gamma_{\,\rm ns}^{\,-,\rm v}(N=3)$ 
require an effort comparable to that for the N$^4$LO corrections 
for Higgs decay to hadrons in the heavy top-quark limit 
\cite{Hgg4}, the hardest calculation performed before with the 
R$^{\ast\!}$ program of ref.~\cite{HR-Rstar}. A full extension to
higher $N$ is not realistic with the present setup.

Our notation for the twist-2 anomalous dimensions and their
perturbative expansion is
\beq
\label{gamExp}
  \gamma_{\,\rm ns}^{\;\rm a}(N) 
  \;=\; - P_{\rm ns}^{\:\rm a}(N) 
  \;=\; \sum_{n=0} \gamma_{\,\rm ns}^{\,(n)\,\rm a}(N) \: \ar(n+1)
\quad \mbox{ with } \quad
  \ars \;=\; \frct{\als(\mufs)}{4\pi}  
\:\: .
\eeq
Here and in eqs.~(\ref{gns4pN2}) -- (\ref{gnsmN3}) below we 
identify
the renormalization scale 
$\mu_{r}$ with the factorization scale $\muf$.
The expansion of $\gamma_{\,\rm ns}^{\:+}(N=2)$ to the fourth 
order in $\ars$ and the 4-loop contribution to 
$\gamma_{\,\rm ns}^{\:-,\rm v}(N=3)$ have been written down in 
eqs.~(B.1), (B.9) and (B.16) of ref.~\cite{MRUVV1}, see also 
refs.~\cite{BCPns06,VelizN2,VelizN34,BCKrev15}.
The lower orders of the latter can be found in appendix C of
ref.~\cite{ReteyV} where, however, the normalization of the group 
factor $d_{abc} d^{\,abc}$ is larger by a factor of 16; see the 
discussion below eq.~(30) in ref.~\cite{MVVp2}.

\pagebreak
 
Our new (except for the $\cf \nff$ terms which are identical to 
those obtained already in ref.~\cite{JAG94}) five-loop contributions 
to these anomalous dimensions read
\bea
\label{gns4pN2}
\lefteqn{ \gamma_{\,\rm ns}^{\,(4)+}(N\!=\!2) \;=\; }
\nn \\[1mm] &&
        { \cfi} \,\* \bigg[\,
            \frct{ 9306376 }{ 19683 }
          - \frct{ 802784 }{ 729 }\, \* \z3
          - \frct{ 557440 }{ 81 }\, \* \z5
          + \frct{ 12544 }{ 9 }\, \* \zts
          + 8512\, \* \z7
          \bigg]
\nn \\[0.5mm] &&  \mbox{\hspn}
      - { \ca \* \cff} \, \*  \bigg[\,
            \frct{ 81862744 }{ 19683 }
          - \frct{ 1600592 }{ 243 }\, \* \z3
          + \frct{ 59840 }{ 81 }\, \* \z4
          - \frct{ 142240 }{ 27 }\, \* \z5
          + 3072\, \* \zts
          - \frct{ 35200 }{ 9 }\, \* \z6
          + 19936\, \* \z7
          \bigg]
\nn  \\[0.5mm] && \mbox{\hspn}
       + { \cas \* \cft} \, \*  \bigg[\,
            \frct{ 63340406 }{ 6561 }
          - \frct{ 1003192 }{ 243 }\, \* \z3
          - \frct{ 229472 }{ 81 }\, \* \z4
          + \frct{ 61696 }{ 27 }\, \* \z5
          + \frct{ 30976 }{ 9 }\, \* \zts
          - \frct{ 35200 }{ 9 }\, \* \z6
          + 15680\, \* \z7
          \bigg]
\nn \\[0.5mm] &&  \mbox{\hspn}
       - { \cat \* \cfs} \, \*  \bigg[\,
            \frct{ 220224724 }{ 19683 }
          + \frct{ 4115536 }{ 729 }\, \* \z3
          - \frct{ 170968 }{ 27 }\, \* \z4
          - \frct{ 3640624 }{ 243 }\, \* \z5
          + \frct{ 70400 }{ 27 }\, \* \zts
          + \frct{ 123200 }{ 27 }\, \* \z6
          + \frct{ 331856 }{ 27 }\, \* \z7
          \bigg]
\nn \\[0.5mm] && \mbox{\hspn}
       + { \caf \* \cf} \, \*  \bigg[\,
            \frct{ 266532611 }{ 39366 }
          + \frct{ 2588144 }{ 729 }\, \* \z3
          - \frct{ 221920 }{ 81 }\, \* \z4
          - \frct{ 3102208 }{ 243 }\, \* \z5
          + \frct{ 74912 }{ 81 }\, \* \zts
          + \frct{ 334400 }{ 81 }\, \* \z6
          + \frct{ 178976 }{ 27 }\, \* \z7
          \bigg]
\quad
\nn	\\ &&  \mbox{\hspn}
       - { \dfAAna \* \cf} \, \*  \bigg[\,
            \frct{ 15344 }{ 81 }
          - \frct{ 12064 }{ 27 }\, \* \z3
          - \frct{ 704 }{ 3 }\, \* \z4
          + \frct{ 58400 }{ 27 }\, \* \z5
          - \frct{ 6016 }{ 3 }\, \* \zts
          - \frct{ 19040 }{ 9 }\, \* \z7
          \bigg]
\nn \\ &&  \mbox{\hspn}
       + { \dfFAnc \* \cf} \, \*  \bigg[\,
            \frct{ 23968 }{ 81 }
          - \frct{ 733504 }{ 81 }\, \* \z3
          + \frct{ 176320 }{ 81 }\, \* \z5
          + \frct{ 6400 }{ 3 }\, \* \zts
          + \frct{ 77056 }{ 9 }\, \* \z7
          \bigg]
\nn \\ &&  \mbox{\hspn}
       - { \dfFAnc \* \ca} \, \*  \bigg[\,
            \frct{ 82768 }{ 81 }
          - \frct{ 555520 }{ 81 }\, \* \z3
          + \frct{ 10912 }{ 9 }\, \* \z4
          - \frct{ 1292960 }{ 81 }\, \* \z5
          + \frct{ 84352 }{ 27 }\, \* \zts
          + \frct{ 140800 }{ 27 }\, \* \z6
          + 12768\, \* \z7
          \bigg]
\nn \\[0.5mm] &&  \mbox{\hspn}
       + { \nf\, \* \cff} \, \*  \bigg[\, 
            \frct{ 1824964 }{ 19683 } 
          - \frct{ 463520 }{ 243 }\, \* \z3 
          + \frct{ 21248 }{ 81 }\, \* \z4
          - \frct{ 16480 }{ 81 }\, \* \z5 
          + \frct{ 6656 }{ 9 }\, \* \zts 
          - \frct{ 6400 }{ 9 }\, \* \z6
          + \frct{ 8960 }{ 3 }\, \* \z7 \bigg]
\nn \\[0.5mm] &&  \mbox{\hspn}
       - { \nf\, \* \ca\, \* \cft} \, \*  \bigg[\,  
            \frct{ 3375082 }{ 6561 } 
          - \frct{ 420068 }{ 243 }\, \* \z3 
          - \frct{ 48256 }{ 81 }\, \* \z4 
          + \frct{ 458032 }{ 81 }\, \* \z5 
          + \frct{ 3968 }{ 3 }\, \* \zts 
          - \frct{ 8000 }{ 3 }\, \* \z6 
          + \frct{ 4480 }{ 3 }\, \* \z7 \bigg]
\nn \\[0.5mm] &&  \mbox{\hspn}
       + { \nf\, \* \cas\, \* \cfs} \, \*  \bigg[\,
            \frct{ 15291499 }{ 13122 } 
          + \frct{ 1561600 }{ 243 }\, \* \z3 
          - \frct{ 114536 }{ 27 }\, \* \z4 
          - \frct{ 252544 }{ 243 }\, \* \z5 
          + \frct{ 24896 }{ 27 }\, \* \zts 
          + \frct{ 13600 }{ 27 }\, \* \z6 
          + \frct{ 11200 }{ 27 }\, \* \z7 \bigg]
\nn \\[0.5mm] &&  \mbox{\hspn}
       - { \nf\, \* \cat\, \* \cf} \, \*  \bigg[\,  
            \frct{ 48846580 }{ 19683 } 
          + \frct{ 4314308 }{ 729 }\, \* \z3 
          - \frct{ 274768 }{ 81 }\, \* \z4 
          - \frct{ 1389080 }{ 243 }\, \* \z5 
          + \frct{ 27808 }{ 81 }\, \* \zts 
          + \frct{ 184000 }{ 81 }\, \* \z6 
          + \frct{ 39088 }{ 27 }\, \* \z7 \bigg]
\nn \\ &&  \mbox{\hspn}
       + { \nf\, \* \dfFAnc} \, \*  \bigg[\, 
            \frct{ 22096 }{ 27 } 
          + \frct{ 43712 }{ 81 }\, \* \z3 
          - \frct{ 512 }{ 9 }\, \* \z4 
          - \frct{ 217280 }{ 81 }\, \* \z5 
          - \frct{ 25088 }{ 27 }\, \* \zts 
          + \frct{ 25600 }{ 27 }\, \* \z6 
          - 2464\, \* \z7 \bigg] 
\nn \\ &&  \mbox{\hspn}
       - { \nf\, \* \cf\, \* \dfFFnc} \, \*  \bigg[\, 
            \frct{ 170752 }{ 81 }
          - \frct{ 328832 }{ 81 }\, \* \z3 
          + \frct{ 650240 }{ 81 }\, \* \z5 
          - \frct{ 8192 }{ 9 }\, \* \zts 
          - \frct{ 35840 }{ 9 }\, \* \z7 \bigg]
\nn \\ &&  \mbox{\hspn}
       + { \nf\, \* \ca\, \* \dfFFnc} \, \*  \bigg[\, 
            \frct{ 207824 }{ 81 } 
          + \frct{ 251392 }{ 81 }\, \* \z3 
          - \frct{ 5632 }{ 9 }\, \* \z4 
          - \frct{ 522880 }{ 81 }\, \* \z5 
          + \frct{ 15872 }{ 27 }\, \* \zts 
          + \frct{ 70400 }{ 27 }\, \* \z6 
          - \frct{ 29120 }{ 9 }\, \* \z7 \bigg]
\nn \\[0.5mm] &&  \mbox{\hspn}
       + { \nfs\, \* \cft} \, \*  \bigg[\,
            \frct{ 1082297 }{ 6561 } 
          - \frct{ 145792 }{ 243 }\, \* \z3 
          + \frct{ 1072 }{ 81 }\, \* \z4
          + \frct{ 55552 }{ 81 }\, \* \z5 
          + \frct{ 1792 }{ 9 }\, \* \zts 
          - \frct{ 3200 }{ 9 }\, \* \z6 \bigg]
\nn \\[0.5mm] &&  \mbox{\hspn}
       + { \nfs\, \* \ca\, \* \cfs} \, \*  \bigg[\, 
            \frct{ 332254 }{ 2187 } 
          - \frct{ 85016 }{ 243 }\, \* \z3 
          + \frct{ 20752 }{ 27 }\, \* \z4 
          - \frct{ 28544 }{ 81 }\, \* \z5 
          - \frct{ 13952 }{ 27 }\, \* \zts 
          + \frct{ 1600 }{ 27 }\, \* \z6 \bigg]
\nn \\[0.5mm] &&  \mbox{\hspn}
       + { \nfs\, \* \cas\, \* \cf} \, \*  \bigg[\, 
            \frct{ 631400 }{ 6561 } 
          + \frct{ 214268 }{ 243 }\, \* \z3 
          - 784\, \* \z4
          - \frct{ 53344 }{ 243 }\, \* \z5 
          + \frct{ 25472 }{ 81 }\, \* \zts 
          + \frct{ 22400 }{ 81 }\, \* \z6 \bigg]
\nn \\ &&  \mbox{\hspn}
       - { \nfs\, \* \dfFFnc} \, \*  \bigg[\,  
            \frct{ 43744 }{ 81 } 
          - \frct{ 35648 }{ 81 }\, \* \z3 
          - \frct{ 1792 }{ 9 }\, \* \z4 
          - \frct{ 52480 }{ 81 }\, \* \z5 
          + \frct{ 2048 }{ 27 }\, \* \zts 
          + \frct{ 12800 }{ 27 }\, \* \z6 \bigg]
\nn \\[0.5mm] &&  \mbox{\hspn}
       + { \nft\, \* \cfs} \, \*  \bigg[\, 
            \frct{ 265510 }{ 19683 } 
          + \frct{ 11872 }{ 729 }\, \* \z3 
          - \frct{ 128 }{ 3 }\, \* \z4 
          + \frct{ 512 }{ 27 }\, \* \z5 \bigg]
\nn \\[0.5mm] &&  \mbox{\hspn}
       + { \nft\, \* \ca\, \* \cf} \, \*  \bigg[\,
            \frct{ 168677 }{ 19683 } 
          + \frct{ 11872 }{ 729 }\, \* \z3 
          + \frct{ 2752 }{ 81 }\, \* \z4 
          - \frct{ 4096 }{ 81 }\, \* \z5 \bigg]
       \,-\, { \nff\, \* \cf} \, \*  \bigg[\,  
            \frct{ 5504 }{ 19683 } 
          + \frct{ 1024 }{ 729 }\, \* \z3 
          - \frct{ 128 }{ 81 }\, \* \z4 \bigg]
\; ,
\eea
\bea
\label{gns4mN3}
\lefteqn{ \gamma_{\,\rm ns}^{\,(4)-}(N\!=\!3) \;=\; }
\nn \\[1.0mm] &&
        { \cfi} \,\* \bigg[\,
            \frct{ 81472935625 }{ 80621568 }
          + \frct{ 99382175 }{ 23328 }\, \* \z3
          - \frct{ 3395975 }{ 162 }\, \* \z5
          - \frct{ 9650 }{ 9 }\, \* \zts 
          + \frct{ 34685 }{ 2 }\, \* \z7 \bigg]
\nn \\[0.5mm] &&  \mbox{\hspn}
       - { \ca\, \* \cff}\, \*  \bigg[\,
            \frct{ 286028134219 }{ 80621568 }
          - \frct{ 23916529 }{ 7776 }\, \* \z3
          - 4490\, \* \zts
          + \frct{ 134090 }{ 81 }\, \* \z4
          - \frct{ 2468075 }{ 108 }\, \* \z5
          - \frct{ 55000 }{ 9 }\, \* \z6
          + \frct{ 155155 }{ 4 }\, \* \z7 \bigg]
\nn \\[0.5mm] &&  \mbox{\hspn}
       + { \cas\, \* \cft} \, \*  \bigg[ \,
            \frct{ 20173099267 }{ 3359232 }
          - \frct{ 15401281 }{ 864 }\, \* \z3
          + \frct{ 732787 }{ 1296 }\, \* \z4
          + \frct{ 1972075 }{ 216 }\, \* \z5
          - \frct{ 63830 }{ 9 }\, \* \zts 
          - \frct{ 79750 }{ 9 }\, \* \z6
          + \frct{ 139895 }{ 4 }\, \* \z7 \bigg]
\nn \\[0.5mm] &&  \mbox{\hspn}
       - { \cat\, \* \cfs} \*  \bigg[
            \frct{ 166662991819 }{ 20155392 }
          - \frct{ 36397493 }{ 2916 }\, \* \z3
          - \frct{ 103763 }{ 54 }\, \* \z4
          + \frct{ 30994565 }{ 3888 }\, \* \z5
          - \frct{ 133990 }{ 27 }\, \* \zts
          - \frct{ 72875 }{ 54 }\, \* \z6
          + \frct{ 2127335 }{ 108 }\, \* \z7 \bigg]
\hspace*{-4mm}\nn \\[0.5mm] &&  \mbox{\hspn}
       + { \caf\, \* \cf} \*  \bigg[ 
            \frct{ 75932079965 }{ 10077696 }
          - \frct{ 27693563 }{ 23328 }\, \* \z3
          - \frct{ 1791229 }{ 1296 }\, \* \z4
          - \frct{ 9417425 }{ 1944 }\, \* \z5
          - \frct{ 96700 }{ 81 }\, \* \zts
          + \frct{ 163625 }{ 81 }\, \* \z6
          + \frct{ 199640 }{ 27 }\, \* \z7 \bigg]
\nn \\ &&  \mbox{\hspn}
       - { \dfAAna\, \* \cf} \, \*  \bigg[\,
            \frct{ 81725 }{ 162 }
          - \frct{ 33505 }{ 18 }\, \* \z3
          - \frct{ 1100 }{ 3 }\, \* \z4
          + \frct{ 52025 }{ 18 }\, \* \z5
          - \frct{ 7000 }{ 3 }\, \* \zts
          - \frct{ 48125 }{ 36 }\, \* \z7 \bigg]
\nn \\ &&  \mbox{\hspn}
       - { \dfFAnc\, \* \cf} \, \*  \bigg[\,
            \frct{ 231575 }{ 36 }
          + \frct{ 6351445 }{ 324 }\, \* \z3
          - \frct{ 2927225 }{ 162 }\, \* \z5
          + \frct{ 23210 }{ 3 }\, \* \zts 
          - \frct{ 200410 }{ 9 }\, \* \z7 \bigg]
\nn \\ &&  \mbox{\hspn}
       + { \dfFAnc\, \* \ca} \, \*  \bigg[\, 
            \frct{ 165871 }{ 54 }
          + \frct{ 1816625 }{ 162 }\, \* \z3
          - \frct{ 41800 }{ 9 }\, \* \z4
          - \frct{ 4456145 }{ 162 }\, \* \z5
          + \frct{ 196880 }{ 27 }\, \* \zts 
          + \frct{ 200750 }{ 27 }\, \* \z6
          - \frct{ 7525 }{ 4 }\, \* \z7 \bigg]
\nn \\[0.5mm] &&  \mbox{\hspn}
       + { \nf\, \* \cff} \, \*  \bigg[\,
            \frct{ 1776521549 }{ 40310784 }
          - \frct{ 1332919 }{ 486 }\, \* \z3
          + \frct{ 5000 }{ 9 }\, \* \zts 
          + \frct{ 33290 }{ 81 }\, \* \z4
          - \frct{ 30325 }{ 81 }\, \* \z5
          - \frct{ 10000 }{ 9 }\, \* \z6
          + \frct{ 14000 }{ 3 }\, \* \z7 \bigg]
\nn \\[0.5mm] &&  \mbox{\hspn}
       - { \nf\, \* \ca\, \* \cft} \, \*  \bigg[\,
            \frct{ 3737356319 }{ 3359232 }
          - \frct{ 2327111 }{ 432 }\, \* \z3
          + \frct{ 1280 }{ 3 }\, \* \zts
          + \frct{ 262069 }{ 648 }\, \* \z4
          + \frct{ 1693715 }{ 162 }\, \* \z5
          - \frct{ 14000 }{ 3 }\, \* \z6
          + \frct{ 7000 }{ 3 }\, \* \z7 \bigg]
\nn \\[0.5mm] &&  \mbox{\hspn}
       + { \nf\, \* \cas\, \* \cfs} \, \*  \bigg[\,
            \frct{ 5637513931 }{ 3359232 }
          + \frct{ 2711207 }{ 486 }\, \* \z3
          - \frct{ 5020 }{ 27 }\, \* \zts
          - \frct{ 457499 }{ 108 }\, \* \z4
          + \frct{ 508820 }{ 243 }\, \* \z5
          - \frct{ 20375 }{ 27 }\, \* \z6
          + \frct{ 50155 }{ 108 }\, \* \z7 \bigg]
\nn \\[0.5mm] &&  \mbox{\hspn}
       - { \nf\, \* \cat\, \* \cf} \*  \bigg[
            \frct{ 8766012215 }{ 2519424 }
          + \frct{ 45697231 }{ 5832 }\, \* \z3
          + \frct{ 1195 }{ 81 }\, \* \zts
          - \frct{ 2848403 }{ 648 }\, \* \z4
          - \frct{ 1808870 }{ 243 }\, \* \z5
          + \frct{ 222250 }{ 81 }\, \* \z6
          + \frct{ 250915 }{ 108 }\, \* \z7
          \bigg]
\nn \\ &&  \mbox{\hspn}
       - { \nf\, \* \cf\, \* \dfFFnc} \, \*  \bigg[\,
            \frct{ 24385 }{ 27 }
          - \frct{ 334010 }{ 81 }\, \* \z3
          - \frct{ 8480 }{ 9 }\, \* \zts 
          + \frct{ 1622600 }{ 81 }\, \* \z5
          - \frct{ 135380 }{ 9 }\, \* \z7 \bigg]
\nn \\ &&  \mbox{\hspn}
       + { \nf\, \* \dfFAnc} \, \*  \bigg[\, 
            \frct{ 297889 }{ 162 }
          + \frct{ 154970 }{ 81 }\, \* \z3
          - \frct{ 62600 }{ 27 }\, \* \zts
          + \frct{ 3700 }{ 9 }\, \* \z4
          - \frct{ 122780 }{ 81 }\, \* \z5
          - \frct{ 36500 }{ 27 }\, \* \z6 - 910\, \* \z7 \bigg]
\nn \\ &&  \mbox{\hspn}
       + { \nf\, \* \ca\, \* \dfFFnc} \, \*  \bigg[\, 
            \frct{ 241835 }{ 162 }
          + \frct{ 333487 }{ 81 }\, \* \z3
          + \frct{ 30560 }{ 27 }\, \* \zts 
          - 10780/9\, \* \z4
          - \frct{ 316900 }{ 81 }\, \* \z5
          + \frct{ 110000 }{ 27 }\, \* \z6
          - \frct{ 71960 }{ 9 }\, \* \z7 \bigg]
\nn \\[0.5mm] &&  \mbox{\hspn}
       + { \nfs\, \* \cft} \, \*  \bigg[\, 
            \frct{ 512848319 }{ 1679616 }
          - \frct{ 57109 }{ 54 }\, \* \z3
          + \frct{ 2800 }{ 9 }\, \* \zts
          + \frct{ 9118 }{ 81 }
         \, \* \z4
          + \frct{ 86440 }{ 81 }\, \* \z5
          - \frct{ 5000 }{ 9 }\, \* \z6 \bigg]
\nn \\[0.5mm] &&  \mbox{\hspn}
       + { \nfs\, \* \ca\, \* \cfs} \, \*  \bigg[\,
            \frct{ 1080083 }{ 5832 }
          - \frct{ 296729 }{ 972 }\, \* \z3
          - \frct{ 21800 }{ 27 }\, \* \zts 
          + \frct{ 56327 }{ 54 }\, \* \z4
          - \frct{ 42860 }{ 81 }\, \* \z5
          + \frct{ 2500 }{ 27 }\, \* \z6 \bigg]
\nn \\[0.5mm] &&  \mbox{\hspn}
       + { \nfs\, \* \cas\, \* \cf} \, \*  \bigg[\, 
            \frct{ 61747877 }{ 419904 }
          + \frct{ 2496811 }{ 1944 }\, \* \z3
          + \frct{ 39800 }{ 81 }\, \* \zts 
          - \frct{ 3503 }{ 3 }\, \* \z4
          - \frct{ 88990 }{ 243 }\, \* \z5
          + \frct{ 35000 }{ 81 }\, \* \z6 \bigg]
\nn \\ &&  \mbox{\hspn}
       - { \nfs\, \* \dfFFnc} \, \*  \bigg[\,
            \frct{ 19435 }{ 27 }
          - \frct{ 53366 }{ 81 }\, \* \z3
          + \frct{ 3200 }{ 27 }\, \* \zts
          - \frct{ 3160 }{ 9 }\, \* \z4
          - \frct{ 70000 }{ 81 }\, \* \z5
          + \frct{ 20000 }{ 27 }\, \* \z6 \bigg]
\nn \\[0.5mm] &&  \mbox{\hspn}
       + { \nft\, \* \cfs} \, \*  \bigg[ 
            \frct{ 28758139 }{ 1259712 }
          + \frct{ 21673 }{ 729 }\, \* \z3
          - \frct{ 610 }{ 9 }\, \* \z4
          + \frct{ 800 }{ 27 }\, \* \z5
          \bigg]
\nn \\[0.5mm] &&  \mbox{\hspn}
       + { \nft\, \* \ca\, \* \cf} \, \*  \bigg[\, 
            \frct{ 13729181 }{ 1259712 }
          + \frct{ 14947 }{ 729 }\, \* \z3
          + \frct{ 4390 }{ 81 }\, \* \z4
          - \frct{ 6400 }{ 81 }
         \, \* \z5 \bigg]
       \,-\, { \nff\, \* \cf} \, \*  \bigg[\,
            \frct{ 259993 }{ 629856 }
          + \frct{ 1660 }{ 729 }\, \* \z3
          - \frct{ 200 }{ 81 }\, \* \z4 \bigg]
\eea
and 
\pagebreak
\bea
\label{gns4vN3}
\lefteqn{ \gamma_{\,\rm ns}^{\,(4)\,\rm v}(N\!=\!3) \:\:=\:\:	
          \gamma_{\,\rm ns}^{\,(4) -}(N\!=\!3) 
} \nn \\[1mm] && \mbox{}
       + \nf\, \* \dtFFnc\, \* \bigg\{\,
         \cfs \, \*  \bigg[\,
            \frct{ 79906955 }{ 46656 }
          + \frct{ 246955 }{ 54 }\, \* \z3
          - \frct{ 504550 }{ 81 }\, \* \z5 \bigg]
\nn \\[0.5mm] &&  \mbox{\hspp}
       - \ca\, \* \cf \, \*  \bigg[\,
            \frct{ 9797321 }{ 3888 }
          - \frct{ 475655 }{ 54 }\, \* \z3
          + \frct{ 17600 }{ 9 }\, \* \z4
          + \frct{ 516950 }{ 81 }\, \* \z5
          - \frct{ 500 }{ 9 }\, \* \zts
          + \frct{ 2800 }{ 9 }\, \* \z7 \bigg]
\nn \\[0.5mm] &&  \mbox{\hspp}
       + \cas \, \*  \bigg[\,
            \frct{ 166535 }{ 486 }
          - \frct{ 1783913 }{ 324 }\, \* \z3
          + \frct{ 5555 }{ 9 }\, \* \z4
          + \frct{ 507515 }{ 81 }\, \* \z5
          - \frct{ 2035 }{ 27 }\, \* \zts
          - \frct{ 5500 }{ 27 }\, \* \z6
          - \frct{ 2765 }{ 18 }\, \* \z7 \bigg]
\nn \\[0.5mm] &&  \mbox{\hspp}
      + \nf\, \* \ca \, \*  \bigg[\,
            \frct{ 285985 }{ 3888 }
          + \frct{ 41954 }{ 81 }\, \* \z3
          + \frct{ 160 }{ 27 }\, \* \zts
          - \frct{ 1010 }{ 9 }\, \* \z4
          - \frct{ 56480 }{ 81 }\, \* \z5
          + \frct{ 1000 }{ 27 }\, \* \z6 \bigg]
\nn \\[0.5mm] &&  \mbox{\hspp}
       + \nf\, \* \cf \, \*  \bigg[\,
            \frct{ 1098323 }{ 3888 }
          - \frct{ 49720 }{ 81 }\, \* \z3
          + \frct{ 3200 }{ 9 }\, \* \z4 \bigg]
       - \nfs\, \*  \bigg[\,
            \frct{ 21823 }{ 1944 } \bigg]
       \,\bigg\}
\; .
\eea
Here $N_A$ and $N_F$ are the dimensions of the adjoint and fermion
representation, with $N_A=8$ and $N_F=3$ in QCD, where the quadratic,
cubic and quartic group invariants take the values 
$C_A = 3$, $C_F = 4/3$, $d_{abc} d^{\,abc} = 5/6$ and 
$d^{\,(4)}_{AA} \equiv d_A^{\,abcd}d_A^{\,abcd} = 135$, 
$d^{\,(4)}_{FA} = 15/2$, $d^{\,(4)}_{FF} = 5/12$, 
see ref.~\cite{GrpTh}.

The terms with even-$n$ values of Riemann's $\zeta$-function in 
eqs.~(\ref{gns4pN2}) -- (\ref{gns4vN3}) provide a partial check that 
was not yet known at the time of ref.~\cite{Hgg4}. 
Consistent with the `no-$\pi^2\!$ theorem' for Euclidean physical 
quantities \cite{no-pi2a,no-pi2b,no-pi2c,no-pi2d}, the $\z6$ terms 
cancel when the \MSb\ anomalous dimensions are combined with the 
corresponding coefficient functions \cite{MVV6,MVV10,avLL16} to 
physical evolution kernels for the structure functions $F_2$ at $N=2$ 
and $F_3$ at $N=3$ in deep-inelastic scattering (the required 
transformation can by found in eqs.~(2.7) -- (2.9) of ref.~\cite{NV3}).
The $\z4$ terms are removed by an additional transformation to a 
renormalization scheme in which the N$^4$LO beta function 
\cite{beta4,beta4a,beta4b,beta4c} does not include $\z4$-terms, such 
as the {\sc MiniMOM} scheme in the Landau gauge 
\cite{MiniMOM1,MiniMOM2,4LoopPrp} or the scheme introduced in 
ref.~\cite{BMJamin06}.	

Combining eqs.~(\ref{gns4pN2}) and (\ref{gns4mN3}) with the 
lower-order results leads to the numerical QCD expansions
\bea
\label{gnspN2}
  \gamma_{\,\rm ns}^{\,+}(N\!=\!2,  \nf\!=\!0) &\!\!=\!\!&
    0.2829\, \as() ( 1 \,+\, 1.0187\,\as() \,+\, 1.5307\, \as(2)
  + 2.3617\, \as(3) {  \,+\, 4.520 \,\as(4) } + \,\ldots\, )
\; , \nn \\[-1mm] &\cdots& \nn \\[-1mm]
  \gamma_{\,\rm ns}^{\,+}(N\!=\!2,  \nf\!=\!3) &\!\!=\!\!&
    0.2829\,\as() ( 1 \,+\, 0.8695\,\as() \,+\, 0.7980\, \as(2)
  + 0.9258\,\as(3) {  \,+\, 1.781 \,\as(4) } + \,\ldots\, )
\; , \nn \\
  \gamma_{\,\rm ns}^{\,+}(N\!=\!2,  \nf\!=\!4) &\!\!=\!\!&
    0.2829\: \as() ( 1 \,+\, 0.7987\, \as() \,+\, 0.5451\, \as(2)
  + 0.5215\: \as(3) {  \,+\, 1.223 \,\as(4) } + \,\ldots\, )
\; , \nn \\
  \gamma_{\,\rm ns}^{\,+}(N\!=\!2,  \nf\!=\!5) &\!\!=\!\!&
    0.2829\: \as() ( 1 \,+\, 0.7280\,\as() \,+\, 0.2877\, \as(2)
  + 0.1571\: \as(3) {  \,+\, 0.849 \,\as(4) } + \,\ldots\, )
\;\;
\eea
and
\bea
\label{gnsmN3}
  \gamma_{\,\rm ns}^{\,-}(N\!=\!3,  \nf\!=\!0) &\!\!=\!\!&
    0.4421\, \as() ( 1 \,+\, 1.0153\,\as() \,+\, 1.4190\, \as(2)
  + 2.0954\, \as(3) {  \,+\, 3.954 \,\as(4) } + \,\ldots\, )
\; , \nn \\[-1mm] &\cdots& \nn \\[-1mm]
  \gamma_{\,\rm ns}^{\,-}(N\!=\!3,  \nf\!=\!3) &\!\!=\!\!&
    0.4421 \,\as() ( 1 \,+\, 0.7952\,\as() \,+\, 0.7183\, \as(2)
  + 0.7607\, \as(3) {  \:+\, 1.508 \,\as(4) } + \,\ldots\, )
\; , \nn \\
  \gamma_{\,\rm ns}^{\,-}(N\!=\!3,  \nf\!=\!4) &\!\!=\!\!&
    0.4421 \,\as() ( 1 \,+\, 0.7218\,\as() \,+\, 0.4767\, \as(2)
  + 0.3921\, \as(3) {  \,+\, 1.031 \,\as(4) } + \,\ldots\, )
\; , \nn \\
  \gamma_{\,\rm ns}^{\,-}(N\!=\!3,  \nf\!=\!5) &\!\!=\!\!&
    0.4421 \,\as() ( 1 \,+\, 0.6484\,\as() \,+\, 0.2310\, \as(2)
  + 0.0645\, \as(3) {  \,+\, 0.727 \,\as(4) } + \,\ldots\, )
\;\;
\eea
in powers of the strong coupling constant $\als$.
Here we have included $\nf = 0$ besides the physically relevant 
values, since it provides useful information about the behaviour 
of the perturbation series. The new N$^4$LO coefficients in 
eqs.~(\ref{gnspN2}) and (\ref{gnsmN3}) are larger than one may have 
expected from the N$^2$LO and N$^3$LO contributions.

It is interesting in this context to consider the effect of the quartic
group invariants.
For example, the $\nf = 0$ coefficients in eqs.~(\ref{gnspN2}) and
(\ref{gnsmN3}) at N$^3$LO and N$^4$LO can be decomposed as
\bea
\label{quartN2}
  2.3617 &=& 2.0878 \,+\, 0.1096\, d^{\,(4)}_{FA}/{n_c}
\nn \\
  4.520\phantom{0}
         &=& 3.552\phantom{1}
                    \,-\, 0.0430\, d^{\,(4)}_{FA}/{n_c}
                    \,+\, 0.0510\, d^{\,(4)}_{AA}/{N_a}
\eea
and
\bea
\label{quartN3}
  2.0954 &=& 2.0624 \,+\, 0.0132\, d^{\,(4)}_{FA}/{n_c}
\nn \\
  3.954\phantom{1}
         &=& 3.371\phantom{1}
                    \,-\, 0.0171\, d^{\,(4)}_{FA}/{n_c}
                    \,+\, 0.0371\, d^{\,(4)}_{AA}/{N_a}
\eea
with $d^{\,(4)}_{FA}/{n_c} = 2.5$ and $d^{\,(4)}_{AA}/{N_a} = 16.875$,
 see, e.g., appendix C of ref.~\cite{4LoopPrp}. 
Without the rather large contributions of $d^{\,(4)}_{AA}$, which 
enters $\gamma_{\,\rm ns}^{}$ at N$^4$LO for the first time, the series would look much more 
benign with consecutive ratios of 1.4 -- 1.6 between the N$^4$LO, 
N$^3$LO, N$^2$LO and NLO coefficients.
This~sizeable $d^{\,(4)}_{AA}$ contribution ($\,\sim \ncs + 36\,$) also 
implies that the leading large-$\nc$ contribution provides a less good 
approximation at N$^4$LO, at least for low $N$, than at the previous 
orders.

The generalization of the expansion coefficients in eq.~(\ref{gamExp})
to $L \equiv \ln (\mu_r^2/\mufs) \neq 0$ is given by \cite{NV3}
\bea
  \gns(0)(L) &\!\!=\!& \gns(0)
\; , \nn \\[1mm]
  \gns(1)(L) &\!\!=\! & \gns(1) + \beta_0 L\, \gns(0)
\; , \nn \\[1mm]
  \gns(2)(L) &\!\!=\! & \gns(2) + 2\beta_0 L\, \gns(1)
   + \left( \beta_1 L + \beta_0^2 L^2 \right) \gns(0)
\nn \\[1mm]
  \gns(3)(L) &\!\!=\! & \gns(3) + 3\beta_0 L\, \gns(2)
   + \left( 2\beta_1 L + 3\beta_0^2 L^2 \right) \gns(1)
   + \Big( \beta_2 L + \frct{5}{2}\,\beta_1\beta_0 L^2
   + \beta_0^3 L^3 \Big)\, \gns(0)
\; , \nn \\[1mm]
  \gns(4)(L) &\!\!=\! & \gns(4) + 4\beta_0 L\, \gns(3) 
   + \left( 3\beta_1 L + 6\beta_0^2 L^2 \right) \gns(2)
   + \left( 2 \beta_2 L + 7 \beta_1 \beta_0 L^2 + 4\beta_0^3 L^3 
     \right) \gns(1)
\; , \nn \\[0.5mm] & & \mbox{}
   + \Big( \beta_3 L + 3\beta_2\beta_0 L^2 + \frct{3}{2}\, \beta_1^2
   L^2 + \frct{13}{3}\, \beta_1 \beta_0^2 L^3 + \beta_0^4 L^4 \Big)\,
   \gns(0)
\eea
to N$^4$LO, where we have suppressed the superscript `a' of 
eq.~(\ref{gamExp}).
$\beta_{0,1,2,3}$ are the \MSb\ coefficients of the beta function 
up to N$^3$LO \cite{beta3a,beta3b} with $\beta_0 = 11 - 2/3\:\nf$, 
$\beta_1 = 102 - 38/3\:\nf\,$ etc in QCD.

The numerical impact of the higher-order contributions to the
anomalous dimensions $\gamma_{\,\rm ns}^{\,\pm}$ on the evolution of
the $N=2$ and $N=3$ moments of the respective parton distributions
(\ref{qns}) are illustrated in fig.~1.
At $\als(\mufs) = 0.2$ and $\nf=4$, the N$^4$LO corrections
are about 0.15\% at the default choice $\mu_{r} = \muf$ of the
renormalization scale, roughly half the size of their N$^3$LO
counterparts.
Varying $\mu_{r}^{}$ up and down by a factor of 2 one arrives at a 
band with a full width of about 0.7\%. 
The N$^3$LO and N$^4$LO corrections are about twice as large
at a lower scale with $\nf=3$ and $\als(\mufs) = 0.25$.

\begin{figure}[p]
\vspace{-1mm}
\centerline{\epsfig{file=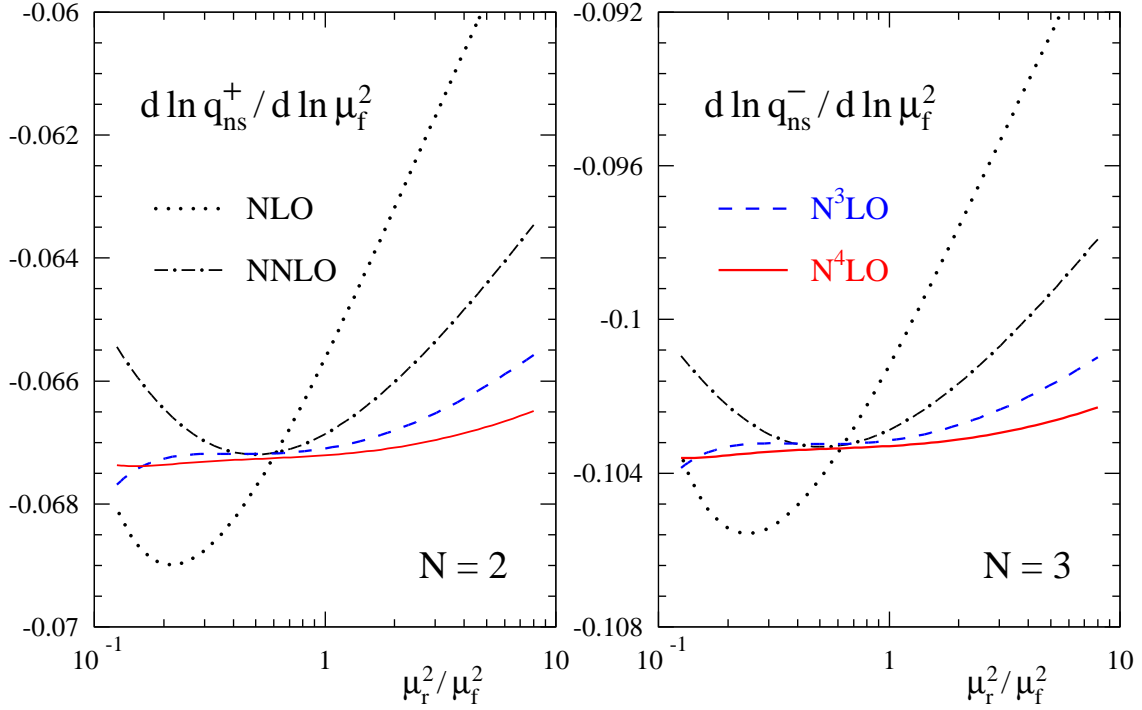,width=15.0cm,angle=0}}
\vspace{-2mm}
\caption{ \label{fig1} \small
The renormalization-scale dependence of the logarithmic 
factorization-scale derivatives of the quark distributions
$q_{\rm ns}^{\,+}$ at $N=2$ and $q_{\rm ns}^{\,-}$ at $N=3$ at a
standard reference point with $\als(\mufs) = 0.2$ and $\nf=4$.
}
\vspace*{-2mm}
\end{figure}

\begin{figure}[p]
\vspace{-1mm}
\centerline{\epsfig{file=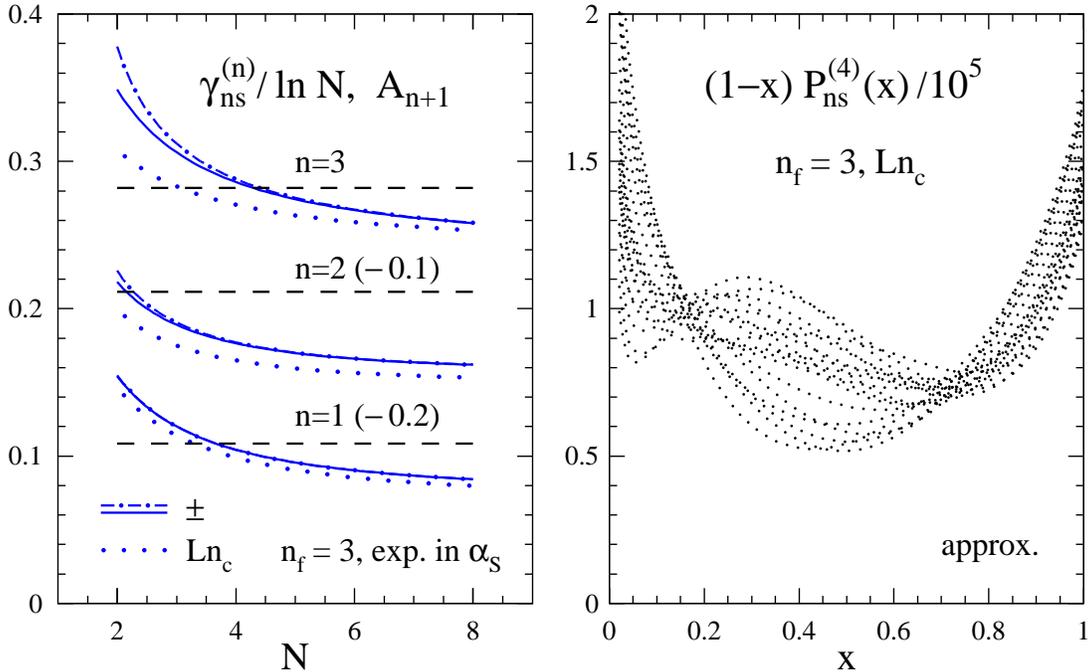,width=15.0cm,angle=0}}
\vspace{-2mm}
\caption{ \label{fig2} \small
Left: non-singlet anomalous dimensions and their generalization 
 to non-even$/$odd $N$ at $2 \leq N \leq 8$. 
 The quantities $\gamma_{\,\rm ns}^{\,(n)\,\pm}(N)/\ln N$ for 
 $\nf = 3$ are compared to their common large-$n_c$ (Ln$_{\,\rm c}$)
 limits and their limits for $N \ra \infty$ (shown as straight 
 lines), the ($n$+1)-loop cusp anomalous dimensions $A_{n+1}$ for 
 $n = 1, 2, 3$.
Right: 20 trial functions incorporating the present integer-$N$
 and endpoint constraints on the 5-loop Ln$_{\,\rm c}$ splitting 
 functions at $\nf = 3$. The resulting uncertainty band for $A_{5}$
 in the large-$n_c$ limit can be read off at $x=1$.
}
\vspace*{-2mm}
\end{figure}

In order to assess the implications of the above results beyond 
$N=2$ and $N=3$, and in particular for the five-loop cusp anomalous 
dimension, it is useful to consider the $N$-dependence of
$\gamma_{\,\rm ns}^{\,\pm}(N)$ at lower orders and the large-$\nc$ 
limit.
In the left part of fig.~\ref{fig2}, moments (\ref{gamExp}) of the 
NLO, N$^2$LO and N$^3$LO splitting functions $P_{\rm ns}^{\,\pm}(x)$ 
and their common large-$n_c$ (Ln$_{\,\rm c}$) limit are displayed for 
$\nf = 3$ in a
manner that facilitates a direct comparison with the size of the 
corresponding cusp anomalous dimensions, defined by 
$A_{\rm q} = A_1^{} \ar() + A_{2\,}^{} \ar(2) + \ldots\,$, 
in eq.~(\ref{Ntoinf}).

At all orders known so far, $\gamma_{\,\rm ns}^{\,(n)\,-}(N)/\ln N$
at $N=3$ deviate from $A_{n+1}$ by less than 8\% for $\nf = 3$. 
The relative deviations are even smaller at $\nf = 0$, but larger at 
larger $\nf$ due to cancellations between the $\nf$-dependent and
$\nf$-independent contributions. However, the corresponding absolute 
deviations at $\nf = 4$ and $\nf = 5$ are comparable to those at 
$\nf = 3$.
 
These results suggest that our above five-loop results can be used 
for a first estimate of the 5-loop cusp anomalous dimension.
The situation is complicated somewhat by the large low-$N$ 
contribution of the new colour structure $d^{\,(4)}_{AA}/N_A$ which 
may or may not persist to the large-$N$ limit. Treating the size of
this contribution as an additional uncertainty, we arrive at the
predictions
\beq
\label{A5estQCD}
  A_5 \:=\: ( 1.7 \,\pm\, 0.5 \:,\; 
              1.1 \,\pm\, 0.5 \:,\;
              0.7 \,\pm\, 0.5 ) \cdot 10^{\,5}
\quad \mbox{for} \quad 
  \nf \:=\: 3\,,\:4\,,\:5
\; .
\eeq
Together with the lower-order results \cite{MVV3,MRUVV1} these lead 
to the QCD expansions
\bea
\label{AqExpQ}
  A_{\rm q}^{}(\nf\!=\!3) &\!\!=\!\!&
  0.42441\:\as() \:
  ( 1  +  0.7266\,\as() +  0.7341\,\as(2) + 0.665\,\as(3)
       +  ( 1.3 \pm 0.4) \as(4) \, + \, \ldots )
\; , \nn \\
  A_{\rm q}^{}(\nf\!=\!4) &\!\!=\!\!& 
  0.42441\:\as() \:
  ( 1  +  0.6382\,\as()  +  0.5100\,\as(2) + 0.317\,\as(3)
       +  ( 0.8 \pm 0.4 ) \as(4)  \, + \, \ldots )
\; , \nn \\
  A_{\rm q}^{}(\nf\!=\!5) &\!\!=\!\!&
  0.42441\:\as() \:
  ( 1  +  0.5497\,\as()  +  0.2840\, \as(2) + 0.013\,\as(3)
       +  (0.5 \pm 0.4 ) \as(4)  \, + \, \ldots )
\quad
\eea
for the physically relevant values of $\nf$. Here and in 
fig.~\ref{fig2} also the N$^3$LO results are approximate;
their uncertainties are however irrelevant and amount to
$2 \cdot\! 10^{\,-4}$ for the coefficients in 
eq.~(\ref{AqExpQ}).

A more direct determination is possible for the leading large-$\nc$ 
contribution of $A_5$.
In this limit $\gamma_{\,\rm ns}^{\,+} = \gamma_{\,\rm ns}^{\,-}$, 
thus the results at $N=1$, $N=2$ and $N=3$ refer to the same function. 
Furthermore, as already noted below eq.~(3.11) in ref.~\cite{MRUVV1},
the large-$\nc$ five-loop coefficients of $C_{\rm q}$ and $D_{\rm q}$ 
in eq.~(\ref{Ntoinf}) can be predicted from known coefficients using
\cite{DMS05,MRUVV1}
\beq
\label{CDofAB}
  C_{\rm q} \;=\; ( \:\! A_{\rm q} \:\! )^{2^{}}
\;\; , \quad
  D_{\rm q} \;=\; A_{\rm q} \cdot ( B_{\rm q}- \beta(\ars)/\ars )
\;\; ,
\eeq
where $\beta(\ars) = - \beta_0\,\ar(2) - \beta_1\,\ar(3) - \ldots\;$.
Finally the coefficients of all Ln$_{\,\rm c}$ small-$x$ logarithms 
at five loops, $\ln^{\,\ell\!} x$ with $\ell = 1, \ldots, 8$, 
can be predicted from the results of ref.~\cite{MRUVV1} 
by solving \cite{Veliz14DL}
\beq
\label{gnsto0V}
  \gamma_{\rm ns}^{}(N,\ars) \cdot 
  \left( \,\gamma_{\rm ns}^{}(N,\ars) + N - \beta(\ars) / \ars 
  \right) \;=\; O(1)
\; \; . 
\eeq
Together, these endpoint constraints imply that the function
$P_{\,\rm ns}^{\,(4)}(x)$ is known in the large-$\nc$ limit up to 
the large-$x$ coefficients $A_5$ and $B_5$ of $1/\xm1_+$ and
$\delta\xm1$, respectively, and a smooth function that approaches 
a constant for $x \ra 0$ and vanishes for $x \ra 1$.
 
Under these circumstances, the three available $N$-values are 
sufficient, just, for a first approximate reconstruction of 
$P_{\,\rm ns}^{\,(4)}(x)$: a sufficient number, here 20, of 
one-parameter smooth functions are chosen, and $A_5$, $B_5$ and
this parameter are determined from the available three moments
for each of these choices. The ensuing spread of the values of
$P_{\,\rm ns}^{\,(4)}(x)$ indicates the remaining uncertainty
of this function. The results are shown in the right part of
fig.~\ref{fig2} for $\nf = 3$ quark flavours.
Corresponding procedures (all of which are, of course, 
mathematically non-rigorous) have been successfully employed to 
three-loop and four-loop quantities in the past, usually with 
(many) more calculated moments but weaker endpoint constraints, 
see, e.g., refs.~\cite{NV4,KlPMV} and ref.~\cite{MRUVV1}.
We have checked the above setup by applying it at N$^3$LO, 
where a comparison with the exact results is possible.

In this manner we arrive at the five-loop Ln$_{\,\rm c}$ 
cusp anomalous dimensions
\beq
\label{A5appLnc}
  A_{5,\,\rm L}  \:=\: 
            ( 1.5  \,\pm\, 0.25 \:,\;
              0.8  \,\pm\, 0.2 \:,\;
              0.4  \,\pm\, 0.1 ) \cdot 10^{\,5}
\quad \mbox{for} \quad
  \nf \:=\: 3\,,\:4\,,\:5
\eeq
and $A_{5,\,\rm L} = ( 4.7 \pm 0.6 ) \cdot 10^{\,5}$ for $\nf = 0$.
Together with the lower-order results, which are here known exactly
to N$^3$LO \cite{MRUVV1,qFFnf1L,qFFnf0L}, these lead to the 
numerical expansions
\bea
\label{AqExpL}
  A_{\rm q,L}^{}(\nf\!=\!3) &\!\!=\!\!&
  0.42441\:\as() \:
  ( 1  +  0.7266\,\as() +  0.7355\,\as(2) + 0.706\,\as(3)
       +  ( 1.1 \pm 0.2)\, \as(4) \, + \, \ldots )
\; , \nn \\
  A_{\rm q,L}^{}(\nf\!=\!4) &\!\!=\!\!&
  0.42441\:\as() \:
  ( 1  +  0.6382\,\as()  +  0.5119\,\as(2) + 0.355\,\as(3)
       +  ( 0.6 \pm 0.2 )\, \as(4)  \, + \, \ldots )
\; , \nn \\
  A_{\rm q,L}^{}(\nf\!=\!5) &\!\!=\!\!&
  0.42441\:\as() \:
  ( 1  +  0.5497\,\as()  +  0.2864\, \as(2) + 0.047\,\as(3)
       +  ( 0.3 \pm 0.1 )\, \as(4)  \, + \, \ldots )
\; . \nn \\ & & 
\eea
These results differ from eq.~(\ref{AqExpQ}) only from the N$^2$LO
contributions which include a (small) term of the form $\cfs\:\!\nf$. 
The largest part to the more sizeable N$^3$LO difference is due to 
the (negative) $d^{\,(4)\!}_{FA}/\nc$ contribution. A difference
between the N$^4$LO QCD and Ln$_{\,\rm c}$ results as shown by the
central values in eqs.~(\ref{AqExpQ}) and (\ref{AqExpL}) would not
be surprising in view of eqs.~(\ref{quartN2}) and (\ref{quartN3}). 
However, the present uncertainties preclude any conclusions even
about the sign of large-$\nc$ suppressed contributions.

Up to N$^2$LO, the gluon cusp anomalous dimension is related to 
its quark counterpart by a simple `Casimir scaling', 
$A_{\rm g} / A_{\rm q} \,=\, C_A/C_F \,=\, 2.25$ in QCD. 
This feature is broken at N$^3$LO by the contributions of the 
quartic group invariants \cite{MRUVV1,BHY17a,BHY17b}, but appears 
to persist in a generalized form to N$^4$LO \cite{Dixon17,MRUVV2} 
that includes the above $C_A/C_F$ relation for the Ln$_{\,\rm c}$ 
contributions.
Assuming that the latter feature holds also at five loops, 
eq.~(\ref{AqExpL}) also provides a first result for the five-loop
gluon cusp anomalous dimension $A_{\rm g,L}$.
If a numerical estimate at N$^4$LO were required of $A_{\rm g}$ in 
QCD, we would recommend, for the time being, to use the last column 
in the main bracket of eq.~(\ref{AqExpL}) with the errors enhanced 
to $\pm 0.6$ (twice the offset between the corresponding 
Ln$_{\,\rm c}$ and full QCD coefficients of $A_{\rm g}$ at N$^3$LO) 
together with the N$^3$LO results in eq.~(4.4) of ref.~\cite{MRUVV2}. 

To summarize, 
we have employed the implementation \cite{HR-Rstar} of the local 
R$^\ast$ operation and the {\sc Forcer} program \cite{Forcer} for 
the parametric reduction of massless self-energy integrals to extend 
previous calculations \cite{BCPns06,VelizN2,VelizN34,BCKrev15,MRUVV1} 
of the anomalous dimensions $\gamma_{\,\rm ns}^{}(N)$ of the 
lowest-$N$ non-singlet \mbox{twist-2} operators to the fifth order 
in the strong coupling constant $\als$.
While the coefficients of $\as(5)$ are larger than expected from 
the lower-order results, these N$^4$LO corrections stabilize the 
numerical results at a sub-percent level; a 1\% correction is 
reached only at $\als = 0.3$ for $\nf = 3$.

At least up to N$^3$LO, the anomalous dimensions 
$\gamma_{\,\rm ns}^{}(N)$ can be written as $f(N) \ln N$, where 
the functions $f$ depend rather weakly on $N$ at $N \geq 3$. 
Assuming that this feature also holds at the present order, our 
results at $N=2$ and $N=3$ set the scale for the N$^4$LO corrections
to the evolution of the non-singlet quark distributions outside the 
small-$x$ region.
Accordingly, we have provided first rough estimates of the large-$N$ 
limit of $\gamma_{\,\rm ns}^{\,(4)}(N)/\ln N$ , the five-loop quark cusp 
anomalous dimension $A_5$, for the physically relevant number of light 
flavours $\nf = 3$, 4 and 5 in QCD. 
A more direct approximate determination of $A_5$ has been presented
in the limit of a large number of colours $\nc$, where the present 
lack of higher-$N$ results is compensated, to a just sufficient 
extent, by constraints on the small-$x$ and large-$x$ limits of the 
corresponding non-singlet splitting functions 
\cite{DMS05,Veliz14DL,MRUVV1}.

A {\sc Form} file with our results in eqs.~(\ref{gns4pN2}) -- 
(\ref{gns4vN3}) and the corresponding lower-order coefficients
can be obtained from the preprint server {\tt https://arXiv.org} 
by downloading the source of this article. It is also available 
from the authors upon request.

%
\subsection*{Acknowledgements}
\vspace*{-1mm}
%

This work has been supported by the Advanced Grant 320651, {\it HEPGAME}, 
of the {\it European Research Council}$\,$ (ERC) and by the {\it Deutsche 
Forschungsgemeinschaft} (DFG) under grant number MO~1801/2-1.
The research of F.~H.\ is supported by the Vidi grant 680-47-551 of the
Netherlands Organisation for Scientific Research (NWO).
For a part of our computations we have used the {\tt ulgqcd} cluster
in Liverpool which was funded by the UK {\it Science \& Technology 
Facilities Council} (STFC) grant ST/H008837/1.

{\footnotesize
\providecommand{\href}[2]{#2}\begingroup\raggedright\endgroup

}

\end{document}